\documentclass[prl,aps,twocolumn, floatfix, superscriptaddress,showpacs]{revtex4}
\usepackage{graphicx,color}
\usepackage{amsmath}
\usepackage{amssymb}
\usepackage{bm}
\begin{document}

%-----title and author----------------------

\title{Brittle and Non-Brittle Events in a Continuum-Granular Earthquake Experiment} 
\author{Drew Geller}
\affiliation{Condensed Matter and Thermal Physics Group}
\author{Robert E. Ecke}
\affiliation{Center for Nonlinear Studies \\ Los Alamos National Laboratory, Los
Alamos, NM 87545}
\author{Karin A. Dahmen}
\affiliation {Department of Physics, University of Illinois Urbana Champaign\\ Urbana, Il 61801}
\author{Scott Backhaus$^1$}
\date{\today}

\begin{abstract}
We report moment distribution results from a laboratory earthquake fault experiment consisting of sheared elastic plates separated by a narrow gap filled with a two dimensional granular medium. Local measurement of strain displacements of the plates at over 800 spatial points located adjacent to the gap allows direct determination of the moments and their spatial and temporal distributions. We show that events consist of localized, larger brittle motions and spatially-extended, smaller non-brittle events.  The non-brittle events have a probability distribution of event moment consistent with an $M^{-3/2}$ power law scaling. Brittle events have a broad, peaked moment distribution and a mean repetition time. As the applied normal force increases, there are more brittle events, and the brittle moment distribution broadens. Our results are consistent with mean field descriptions of statistical models of earthquakes and avalanches. 

\end{abstract}
\maketitle

Earthquakes involve complex stick-slip motion, heterogeneous material properties, and a large range of length scales from less than a meter to many hundreds of kilometers \cite{Ben-ZionRG08}.  One feature of earthquakes is a power law probability distribution of earthquake magnitude known as the Gutenberg-Richter (GR) law.  In terms of the moment $M$ released and for globally averaged strike-slip faults (surfaces moving past each other horizontally) the GR distribution is consistent with $P(M) \sim M^{-1-\beta}$ with $\beta = 1/2$ 
%(equivalent to $b = 1.5 \beta = 0.75$ for magnitude distribution) 
\cite{FrohlichJGeoRes93}.  On individual earthquake faults there are examples of both GR distributions and ones where there is a deficit of smaller events and an excess of larger events \cite{StirlingGeoJInt96}.  Because physical measurements that capture the complexity of real earthquake faults are difficult, theoretical models, laboratory experiments and numerical simulations have been developed to provide insight into earthquake physics. The simplest ingredients of a strike-slip fault model \cite{CarlsonRMP94,Ben-ZionRG08,KawamuraRMP12} are continuum elastic plates representing the differential motion of large-scale tectonic plates with coupling between the plates determined by frictional interactions.  An apparent feature of mature faults is that there is fault gouge - ground up granular matter - at the interface of the plate surfaces \cite{ChesterTect1998}. This feature has prompted consideration of the role of idealized granular media subject to compaction and shear in producing stick-slip fluctuations similar in nature to earthquake events \cite{HowellPRL99,BocquetPRE01,vanHeckeNature03,MajmudarNature05,DanielsJGRSE08,HaymanPAG11}.  Another class of laboratory experiments has focused on effective friction behavior in seismically relevant materials   \cite{MaroneAREPS98} or in block-on-block frictional dynamics \cite{Ben-DavidScience2010}. 

Theoretical models incorporate various ingredients from continuum properties \cite{Ben-ZionRG08} to statistical physics approaches \cite{CarlsonRMP94,KawamuraRMP12} that include discrete granular interactions. These models are usually studied using numerical simulations \cite{Ben-ZionRG08,CarlsonRMP94,RundleROG03,KawamuraRMP12,CiamarraPRL10} but some models are amenable to theoretical analysis and predict universal scaling behavior in the limit of long-range forces, i.e., in the mean field limit \cite{FisherPRL97,DahmenNatPhys11}.  Universal models are compelling in that they do not rely on detailed modeling of all the relevant physics of earthquakes and other stick-slip systems.  The major predictions for mean field theory models \cite{FisherPRL97,DahmenPRL09,DahmenNatPhys11} concern the slip moment $M$ which is the sum of the individual local displacements $s$ induced in an event. These local displacements are spatially distributed along the slip plane as in real earthquakes. The probability distribution $P(M)$ is expected to scale as $M^{-3/2}$ for small $M$ whereas there may be enhanced probability for large moment events if there is ``weakening'' in the effective friction (lower effective friction after an event). 
\begin{figure}[h]
%\vspace{4in}
\includegraphics[width = 2.8 in]{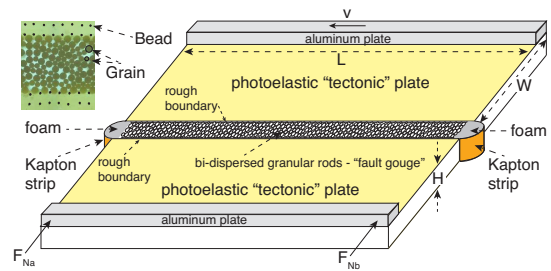}
\caption{(Color online) Schematic illustration of experimental laboratory earthquake apparatus}
\label{fig:apparatus}
\end{figure}

We have developed a laboratory analog of a single strike-slip fault with continuum plates and granular fault gouge.  We measure the local slip $s$ and the total slip moment $M$ associated with stick-slip events that occur as the system is slowly sheared.  The spatial structure of the slip events along this  fault is directly obtained in contrast to other experiments of this class \cite{MaroneAREPS98,HaymanPAG11}.  The experiment has two soft plates, one fixed and the other sheared at fixed velocity, and the forces between the plates are coupled across a narrow gap by a confined quasi two-dimensional granular material.  We find that the total probability distribution of $M$ is consistent with the mean-field prediction \cite{FisherPRL97,DahmenNatPhys11} of $M^{-3/2}$ for small $M$ but shows a crossover to a surplus of large moment events as the applied normal force $F_N$ is increased.  When the events are divided into ``brittle'' (B), large-scale, localized ones and smaller, less localized, ``non-brittle'' (NB) events (defined in detail below), the NB probability distribution is well described by a $M^{-3/2}$ distribution over about two decades in $M$, consistent with predictions of mean field theory \cite{FisherPRL97,DahmenPRL09}. The probability distribution of B events is a broad peaked distribution whose peak shifts towards higher moment and broadens with increasing normal force.  Our measurements lend strong support to the applicability of mean field models to complex stick-slip motion associated with natural earthquake behavior.

Our experimental apparatus, shown in Fig.\ \ref{fig:apparatus}, consists of two plates formed from photo-elastic material (Vishay PS-4) with dimensions $L = 50$ cm, $W = 25$ cm, and $H =  1$ cm. The plate material has an elastic modulus $E_p = 4$ MPa and a Poisson ratio of 0.50. Aluminum supports on the outer edges provide structural support for the flexible plates and are respectively sheared at one boundary using a linear screw driven by a stepper motor and compressed by a normal force $F_N$ in the range $20 < F_N < 200$ N applied to the fixed boundary ($F_N/2$ at opposite ends of the fixed boundary using micrometers and force sensors) at the start of the run.  The plates are spaced $W_g \approx 1$ cm apart and the gap is filled with a bi-dispersed set of 3000 nylon rods with height H and diameters 0.119 and 0.159 cm.  The elastic modulus of the granular rods is $E_g \approx 4$ GPa, 1000 times stiffer than the plate material.  At both ends of the gap, a flexible strip of Kapton sheet confines the granular material and a piece of compressible foam keeps the force more or less constant as the plates translate. The total physical displacement in an experiment is $\pm 2$ cm corresponding to total strain of about 0.04 ($\Delta x/L$), and the shearing velocity $v \approx 4 \mu$m/s corresponds to a strain rate $\dot \gamma_H = v/H = 4 \times 10^{-4}$/s (or more traditionally based on $L$:  $\dot \gamma_L = v/L = 8 \times 10^{-6}$/s).  The strain rate gives a natural time scale $\tau_H = 1/\dot \gamma_H$. The boundary conditions between the elastic plates and the granular material are set by a regularly spaced array of 202 large diameter pins glued into each of the elastic plates at the edge of the fault. Additionally, 812 (4 rows of $N_p$ = 203) ball bearings of 300 $\mu$m diameter are arranged in two rows on each plate near the inner edge with nominal $\ell =$ 0.22 cm separation, as can be seen in the detailed view in the upper left hand corner of Fig. \ref{fig:apparatus}.  Photographic digital images (Olympus E-620: 3024 $\times$ 4032 pixels) of the ball bearings are taken every $\delta t = $ 2.5 sec (500 motor steps) and provide individual locations $\{x_{i},y_{i}\}^j_k$ with relative error of $12\ \mu$m (0.2 pixels) where $i$ labels the distance along the gap, $j$ is the time index in units of $\delta t$, and $k$ indicates the row.  As the plates are sheared, the bead arrays deform to follow the elastic response of the plates. The differential motion between successive time steps of bead $i$ is defined as $s_i^j = x_i^j - x_i^{j-1}$ with values less than the average noise threshold of 0.3 pixels set to zero.  The spatially-integrated moment at time step $j$ is $m^j = \sum_{i=1}^{N_p} s_i^j$. We then convert this to a dimensional moment by multiplying by the area of an event which is the spacing between beads $l$ times the plate height $H$. Finally, we obtain a dimensionless moment for event $j$ of $M^j = m^j \alpha \ell H/(H^2 \ell) = m^j \alpha/H$ where $\alpha = 0.006$ cm/pixel. 
\begin{figure}[h]
%\vspace{4in}
\includegraphics[width = 2.8 in]{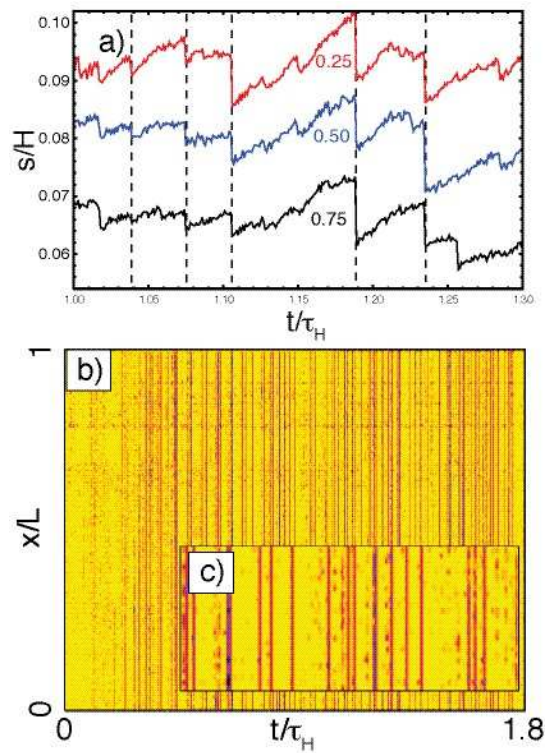}
\caption{(Color online) 
 (a) Typical $s_i/H$ vs $t/\tau_H$ for $x/L$ = 1/4, 1/2, and 3/4 for $F_N = 60$ N. (b) The space-time displacements of beads indicated by intensity where darker points indicate more motion for $F_N = 80$ N.  The ordinate is $x/L$ and normalized time $t/\tau_H$ proceeds from left to right.  (c)  Expanded segment of space-time plot in b).
}
\label{fig:SpaceTime}
\end{figure} 

A critical aspect of our experiments is the ability to determine the spatial distribution of stick-slip events.  In Fig.\ \ref{fig:SpaceTime} (a), we show a typical set of displacements $s_i$ versus normalized time $t/\tau_H$ for lateral positions $x/L$ = 1/4, 1/2, and 3/4. Large displacements indicated by dashed lines occur over the whole fault length whereas the spatial coherence of smaller events varies.  A more indicative representation of spatial coherence can be seen in a space-time plot of the magnitude of the differential displacements in Fig.\ \ref{fig:SpaceTime}(b).  Large, fault-spanning events (solid vertical lines spanning most of $L$ in Figs.\ \ref{fig:SpaceTime}(b),(c))  occur frequently for these conditions (the frequency increases and the events become more nearly periodic at higher normal load).  A close-up view of the motions near the center of the fault in Fig.\ \ref{fig:SpaceTime}(c), shows detail of smaller events in which only a few locations or clusters of beads slip in a given time interval.  In addition to global events and spatially extended ones, there are localized events that have length less than $L$. This variation arises from the heterogeneous force distribution in the granular material set up by stress chains.  To extract more detailed spatial information, the center for an event at time $j$ is defined by the displacement-weighted bead position $X^j = \left( \sum_{i=1}^{N_p} s_i^j x_i^0 \right)/M^j$, where $x_i^0$ is the nominal initial position of the bead $i$.  The degree of spatial localization of an event at time $j$ is then defined by the radius of gyration, $R^j = \left( \sum_{i=1}^{N_p} s_i^j \left( X^j - x_i^0 \right)^2 / M^j\right )^{1/2}$. We also compute a normalized quantity that reflects the spatial extent of an event: $C^j = R^j/\sqrt{12} N_j$ where $N_j$ is the number of non-zero values of $s_i^j$ (the numerical factor gives $C = 1$ for a spatially uniform event of any size with constant amplitude).
\begin{figure}[h]
%\vspace{4in}
\includegraphics[width = 2.8 in]{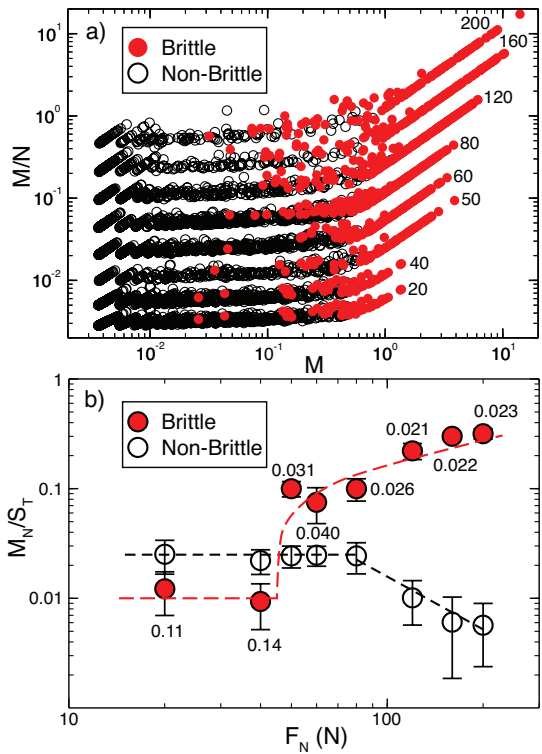}
\caption{(Color online) (a) Average slip M/N vs. M for labeled $F_N$. Curves are offset vertically for clarity (brittle: solid, red - non-brittle: open, black). (b) Fraction of moment $M_N$ relative to available total strain moment $S_T$ (brittle: solid, red;  non-brittle: open, black). Dashed lines - see text. The mean repeat times for B events are labeled.}
\label{fig:slip}
\end{figure}

For events that involve motion at only a small number of locations, the average motion per bead, $M_j/N_j$ in the event is nearly constant as a function of $M$, as seen in Fig.\ \ref{fig:slip}(a).  The average motion typically spans a small range, so that events differ in moment owing to the number of beads involved rather than the distance that they move because an isolated site can only move so far before the elastic reaction from its nearest neighbors significantly increases.  For  high moments, $M/N$ is linear in $M$ because the number of beads participating in a slip event is limited by the size of our experiment, i.e., $N_j \approx N$ so that $M/N$ can only grow in proportion to the average bead displacement (a similar trend is seen for very small $M$). These events may achieve much higher average displacements per bead than the localized events because there are no locations remaining pinned by force chains across the gouge and therefore no elastic reaction.  This behavior is similar to brittle fracture in solids \cite{AlavaAP06} and to the ``characteristic'' events seen in mean field models \cite{FisherPRL97,DahmenPRL09}, and it suggests that events should be divided into B and NB populations.  One way to divide these events is to define B as system spanning events and NB as smaller events (see also \cite{HaymanPAG11}).  We refine this division by including as brittle events localized ones with $C^j  < 2$ and where $M^j/N^j > 0.003$ (twice our noise threshold).  Our results do not depend sensitively on either cutoff. Both types of events are indicated in Fig.\ \ref{fig:slip}(a).

In any particular shear experiment, the system reaches a steady state, i.e., the plastic yield limit, in which the fixed plate no longer has a time-averaged motion accumulating elastic energy.  The shear imparted by the moving plate must be released over time, and we can get a sense of the fluid/solid character of the granular medium, i.e., sliding versus stick-slip motion, by comparing the total event moment $M$ per site $i$, i.e., $M_N = (1/N_p) \sum_{j,k} M_k^j$ with the total displacement of the moving plate $S_T = \sum_k v T_k$ ($T_k$ is the total time of run $k$ in which events are recorded).  Figure \ref{fig:slip}(b) shows that the $M_N$ of all B and NB events recorded for the fixed plate captures less than 30\% of the continual motion of the driven plate over the full range of normal loading $F_N$ studied.  The contribution of the B events increases dramatically in the range $F_N > 45$ N consistent with $(F_N - 45)^{1/2}$ (long dashed curve in Fig.\ \ref{fig:slip}(b) for $F_N > 45$ N), perhaps indicative of a bifurcation, and the contribution of these events is larger than that for NB events except at very low $F_N$.  The NB events account for a constant fraction of the motion up to about 80 N, above which their contribution declines rapidly.  The remaining ($> 70\%$) fraction of the available motion results from steady sliding (contributions from small events below our noise threshold are estimated to be small). In MF models of stick-slip behavior \cite{FisherPRL97,LeBlancPRE13}, sliding does not affect the expected scalings.
\begin{figure}[h]
%\vspace{4in}
\includegraphics[width = 2.8 in]{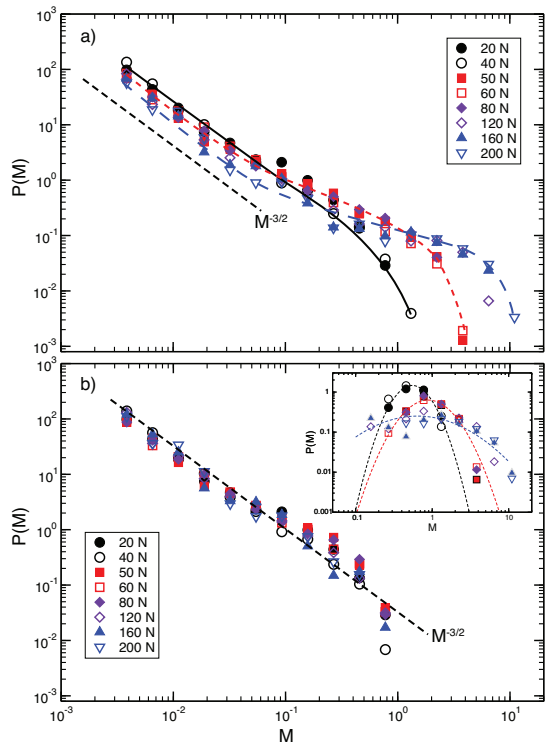}
\caption{(Color online)$P(M)$ vs $M$ for different $F_N$:  (a) all events;  (b) NB events; (c)  B events.  Solid, short-dashed and long-dashed curves show variations with $F_N$ in (b) and (c). $M^{-3/2}$ scaling labeled in (a) and (b).}
\label{fig:moment}
\end{figure}

The probability distribution of all recorded event moments for different $20 < F_N < 200$ N is shown in Fig.\ \ref{fig:moment}(a).  As $F_N$ increases, the probability of NB events decreases, consistent with trends in models \cite{DahmenNatPhys11} that include the effects of packing fraction $\phi$ ($\phi$ increases with increasing $F_N$). The large B events become more probable, growing at the expense of the non-brittle events, and have higher mean $M$ for larger $F_N$. Using our decomposition in B and NB events, we can represent their individual contributions to the overall probability distribution $P(M)$.  In Fig.\ \ref{fig:moment} (b), we show the NB distribution (normalized to the number of NB events) and in the inset Fig.\ \ref{fig:moment} (b) we show the B distribution and its variation with $F_N$.  This decomposition demonstrates cleanly that NB events are distributed as a power law over about 2 decades in M consistent with $M^{-3/2}$ (the remaining depletion/enhancement of $M$ may be a finite size effect).  On the other hand, the B events are concentrated at large $M$ with broader distribution and higher mean $M$ as $F_N$ increases.  (The weakening parameter for our system \cite{DahmenNatPhys11} is of order 0.07 as indicated by the average fractional shear stress drop for the largest B events.)  

In addition to the emergence of excess large-event probability at higher $F_N$, the B events develop a dominant mean repetition time $\tau$ for large $F_N$, see Fig.\ \ref{fig:SpaceTime}(b),(c).  The distributions for low $F_N < 50$ are consistent with an exponential distribution in $\tau$ which is representative of a random Poisson process. For $F_N > 50$, the mean repetition time is $0.02 < \tau/\tau_H < 0.04$ with small overall variation with $F_N$ (values labeled for data points in Fig.\ \ref{fig:slip}(b)).  Many laboratory based experiments and model simulations show the emergence of a mean repetition time associated with large brittle events \cite{Ben-ZionRG08,FisherPRL97,DahmenNatPhys11}. 

The moment distribution scaling for NB events with the emergence of B events combined with the concurrent development of a mean recurrence time $\tau/\tau_H \approx 0.025$ and the demonstrated spatial coherence of B events with increasing $F_N$ gives a cohesive picture of system behavior.  At small normal force, spatial and temporal correlations are weak giving rise to random spatially extended events and power law $M^{-3/2}$ scaling corresponding to the ``fluid" like phase in mean field theory \cite{DahmenNatPhys11}.  As $F_N$ increases, the granular media compacts owing to compressive stresses and more effectively couples motion on either side of the gap.  This coupling leads to a larger fractional moment slipped, a mean recurrence time between events, and many more spatially compact events with $R^j \sim L$. Our results are consistent with a system with frictional weakening with weakening parameter $\epsilon \approx 0.07$.  The full physical picture of the phenomena we report is complex involving the jamming properties, i.e., rheology, of the sheared granular medium. Many other features of the experimental data are possible including a ``microscopic'' elucidation of the individual motions of the grains and their individual or collective motion in events \cite{DanielsJGRSE08}.  

%Acknowledgements: 
We acknowledge conversations with Karen Daniels and Jonathan Uhl. Work at Los Alamos National Laboratory was funded by the National Nuclear Security Administration of the U.S. Department of Energy under Contract No. DE-AC52-06NA25396. K.D. acknowledges support from grants NSF DMR 1005209 and NSF DMS 1069224.
%%%%%%%%%%%%%%%%%%%%%%%%%%%%%%%%%%%%%%%%%%%%%%%%%%%%%%%%%%%%%%%%%%%%%%%%%%%%

\end{document}